\documentclass[letter, traditabstract]{aa} % for the letters 
\usepackage{graphicx}
\usepackage{txfonts}
\usepackage{natbib}

\newcommand{\bell}{$B_{\ell}$}

\def\gtrsim{\mathrel{\hbox{\rlap{\hbox{\lower4pt\hbox{$\sim$}}}\hbox{$>$}}}}
\def\ltsim{\mathrel{\hbox{\rlap{\hbox{\lower4pt\hbox{$\sim$}}}\hbox{$<$}}}}

\begin{document}

\titlerunning{The dramatic change of the fossil magnetic field of HD 190073}
\title{The dramatic change of the fossil magnetic field of HD~190073: evidence of the birth of the convective core in a Herbig star ?
\thanks{Based on observations collected at the Canada-France-Hawaii Telescope (CFHT) which is operated by the National Research Council of Canada, the Institut National des Sciences de l'Univers (INSU) of the Centre National de la Recherche Scientifique (CNRS) of France and the University of Hawaii, at the Observatoire du Pic du Midi (France), operated by INSU, and at the European Southern Observatory, Chile (Program ID 187.D-0917)}
}

\author{E. Alecian \inst{1} \and
            C. Neiner \inst{1} \and
            S. Mathis \inst{2,1} \and
            C. Catala \inst{1} \and
            O. Kochukhov \inst{3} \and
            J. Landstreet \inst{4,5} \and
            the MiMeS collaboration
          }

\institute{LESIA-Observatoire de Paris, CNRS, UPMC Univ., Univ. Paris-Diderot, 5 place Jules Janssen, F-92195 Meudon Principal Cedex, France,
              \email{evelyne.alecian@obspm.fr}
              \and
              Laboratoire AIM, CEA/DSM, CNRS, Universit\'e Paris Diderot, IRFU/SAp Centre de Saclay, F-91191 Gif-sur-Yvette, France
              \and
              Department of Physics and Astronomy, Uppsala University, Box 516, SE-751 20 Uppsala, Sweden
              \and
              Armagh Observatory, College Hill, Armagh, BT61 9DG, Northern Ireland, UK
              \and
              Department of Physics and Astronomy, The University of Western Ontario, London, Ontario, N6A 3K7, Canada
              }

   \date{Received September 15, 1996; accepted March 16, 1997}

\abstract{{ In the context of the ESPaDOnS and Narval spectropolarimetric surveys of Herbig Ae/Be stars, we discovered and then monitored the magnetic field of HD 190073 over more than four years, from 2004 to 2009. Our observations all displayed similar Zeeman signatures in the Stokes $V$ spectra, indicating that HD 190073 hosted an aligned dipole, stable over many years, consistent with a fossil origin. We obtained new observations of the star in 2011 and 2012 and detected clear variations of the Zeeman signature on timescales of days to weeks, indicating that the configuration of its field has changed between 2009 and 2011. Such a sudden change of external structure of a fossil field has never previously been observed in any intermediate or high-mass star. HD~190073 is an almost entirely radiative pre-main sequence star, probably hosting a growing convective core. We propose that this dramatic change is the result of the interaction between the fossil field and the ignition of a dynamo field generated in the newly-born convective core.}}


   \maketitle
%
%________________________________________________________________

\section{Introduction}

A few percent of main-sequence and pre-main sequence (MS) stars of intermediate mass host strong magnetic fields, of simple (topologically dipolar) configuration, stable over many years or decades \citep[e.g.][]{donati09}. Recent studies of Herbig Ae/Be stars \citep{alecian12} as well as of giants \citep{auriere08,auriere09} have demonstrated a fossil link between the magnetic stars at different evolutionary stages: from the pre-MS to the post-MS. The physical processes at the origin of these fields are starting to be understood. Due to the presence of these fields in very young stars, and to the lack of characteristics that could be linked to a dynamo, these fields are believed either to be remnants of the Galactic fields that would have been swept up by the contracting core, or to have been generated by a dynamo when part of the star was convective during the very early phases of its life. This is the fossil theory. { \citet{braithwaite06} have performed numerical simulations showing that an arbitrary initial field quickly evolves into a relaxed stable configuration \citep[see also][]{duez10a,duez10b} and, once formed, evolve on timescales longer than the MS lifetime of a star.} The resulting configurations are similar to those observed on the surface of the intermediate-mass stars. One of the most poorly understood aspects of the fossil scenario is the interplay between the dynamo magnetic fields arising in the convective envelope at the beginning of the pre-MS phase, and in the convective core at the end of the pre-MS, and the fossil magnetic field residing inside the radiative layers of the star.

HD 190073 is an Herbig Ae star with an effective temperature of 9250~K \citep{acke04b}. Its magnetic field was discovered in 2005 as part of a survey of magnetic fields in Herbig Ae/Be stars \citep{catala07}. We first observed it for 4 consecutive days, then, once every few months for more than four years, with ESPaDOnS or Narval, to determine the stability of the field over time. In contrast to most magnetic $V$ signatures observed in intermediate and high-mass stars, the Zeeman signature of HD 190073 was always similar from one observation to another. No variability was detected. In other stars the observed variations are explained by the oblique rotator model, describing a rotating star hosting an inclined dipolar magnetic field \citep{stibbs50}. \citet{catala07} proposed three scenarios to explain the absence of rotational modulation in HD 190073: (i) the star is seen pole-on, (ii) the magnetic and rotation axes are aligned, or (iii) the rotation period of the star is very long.

\begin{figure*}
\centering
\includegraphics[height=\textwidth,clip=true,angle=90]{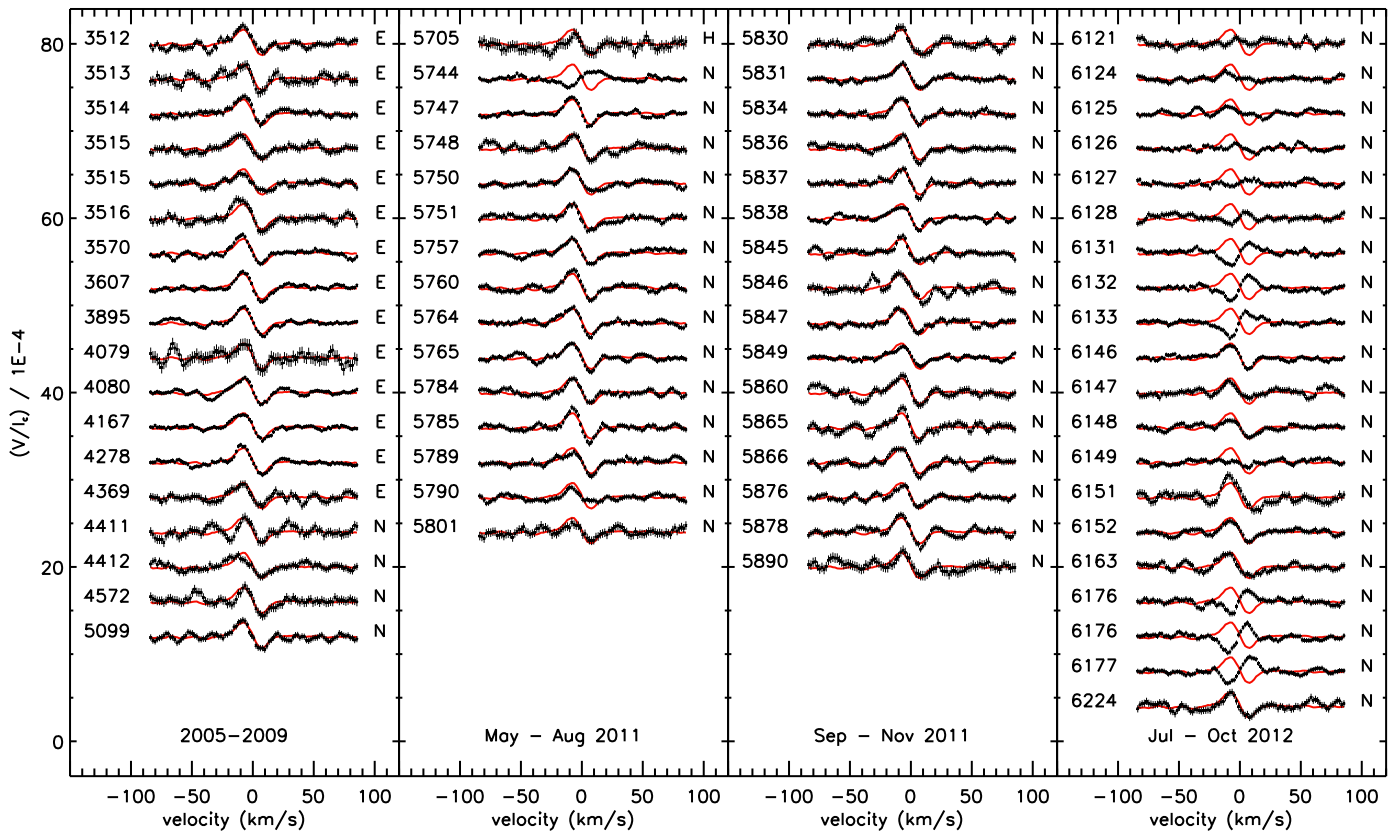}
\caption{LSD $V$ profiles (black) of HD 190073 ordered chronologically from the top left to bottom right. The numbers are the integer part of the HJD. Both 6176 profiles were obtained on the same day, 3.4 h apart. An error bar is superimposed on each pixel. The red profile overplotted on each observation is the average profile for the 2005-2009 data set. The letters refer to the instrument (E: ESPaDOnS, N: Narval, H: HARPSpol).}
\label{fig:var}%
\end{figure*}

The stability of magnetic fields in Herbig Ae/Be stars, including HD 190073, is well established, and is a very strong argument in favour of the fossil theory \citep{braithwaite06}. However, in 2011 and 2012 we obtained a new series of observations (about two years after the end of the first series), using HARPSpol and Narval, and to our amazement we detected strong variations on timescales of few days or weeks, clearly indicating an abrupt change of the magnetic configuration at the surface of HD 190073. This phenomenon has never been observed in any magnetic intermediate or high mass star. In this letter we report the observations (Sect. 2), describe the analysis of the polarised spectra (Sect. 3), and discuss the possible origins of this unprecedented phenomenon (Sect. 4).

%
%__________________________________________________________________

\section{Observations}

We have obtained two sets of observations. The first consists of 18 observations that were acquired between May 2005 and September 2009 with the high-resolution spectropolarimeters ESPaDOnS and Narval, installed at the Canada-France-Hawaii Telescope (CFHT, Hawaii), and at the T\'elescope Bernad Lyot (Pic du Midi, France), respectively. These observations were acquired in the course of PI programs dedicated to the study of the configuration and stability of the magnetic fields in Herbig Ae/Be stars. Some of these observations have already been published in \citet{catala07}. The second set of high-resolution spectropolarimetric observations consists of one HARPSpol (3.6m-ESO Telescope, Chile) spectrum obtained in May 2011, and 46 Narval spectra acquired between July 2011 and August 2012. 
%These observations have been obtained thanks to Large Programs that the MiMeS (Magnetism in Massive Stars) collaboration obtained on both telescopes to establish the magnetic properties of massive stars at different evolutionary stages \citep{wade09}.
These observations were obtained in the framework of the MiMeS (Magnetism in Massive Stars) Large Programs, which are aimed at establishing the magnetic properties of massive stars at different evolutionary stages \citep{wade09}.

All ESPaDOnS and Narval data were reduced as described by \citet{catala07}, using the dedicated tool Libre ESpRIT \citep{donati06}. The HARPSpol observation was reduced using a modified version of the REDUCE package of \citet{piskunov02}, as described by \citet{neiner12}. In order to increase the signal-to-noise ratio (SNR) of our data we used the Least-Squares Deconvolution method \citep[][LSD]{donati97} with the complete Kurucz mask described in \citet{catala07}, i.e. the mask containing almost all lines predicted by the model. This method provides us with the intensity $I$ and circularly polarised Stokes $V$ mean profiles of the spectra. These $V$ profiles are plotted in Fig. \ref{fig:var}. Examples of LSD $I$ profiles can be found in \citet{catala07}. The SNR in the LSD $V$ profiles vary from 7000 to 22\,000 in the two data sets. The longitudinal magnetic field strengths (\bell) were computed as described in \citet{alecian12}, and vary between $-35$~G and $+55$~G.

The intensity spectra and LSD profiles of the second data set have similar characteristics than those of the first set, already described by \citet{catala07}. In both sets the spectrum is strongly contaminated with circumstellar emission, and this emission changes with time in a similar way. The photospheric ($I$) spectral lines of the two data sets have the same shapes, depths, and weak broadening. Unlike the intensity spectra, the circularly polarised spectra are quite different between the two data sets, as detailed and discussed below.

%
%__________________________________________________________________

\section{The variability of the polarised spectra}

In Fig. \ref{fig:var} we plot (in black) the Stokes $V$ LSD profiles obtained between 2005 and 2012. We have overplotted on each profile the average of all profiles obtained between 2005 and 2009 (in red), i.e. all profiles of the first observing set, during which no significant variations were observed. It is clear from Fig. \ref{fig:var} that all profiles of the first set are indeed similar to the average. However, in 2011 and 2012, many profiles look different: sometimes the signature is smaller in amplitude (e.g. HJD 2455753 and 2455790), sometimes there is no signal (e.g. HJD 2456127 and 2456149), and sometimes the sign of the signature is opposite (e.g. HDJ 2455744, the 2456128--2456133 and the 2456176--2456177 series).

\begin{figure}
\centering
\includegraphics[width=5cm,clip=true]{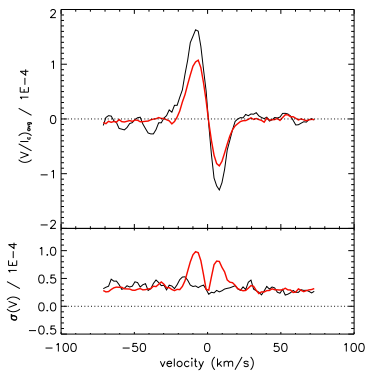}
\caption{Average (top) and standard deviation (bottom) of the LSD $V$ profiles of the first (black, 2005-2009) and second (red, 2011-2012) data sets.}
\label{fig:sig}%
\end{figure}

In order to search for possible significant variation of the LSD $V$ profiles in the first data set, we have computed the standard deviation of the individual $V$ profiles. The results are plotted in Fig. \ref{fig:sig} (in black). If small but significant differences exist between the profiles of the sample, we would expect an increase of the standard deviation inside the profile with respect to the continuum where the signal is null. In Fig. \ref{fig:sig}, we observe that the standard deviation in the black profile is totally flat, which means that no significant variation is detected in the magnetic signature of HD~190073 over more than four years (2005-2009).

The average and standard deviation of the $V$ profiles of the second data set are overplotted in Fig. \ref{fig:sig} (in red). In contrast to the first set, we observe an increase of the standard deviation inside the profile, well above the noise level, indicating that the profiles are not all similar and that the variations were definitely significant in 2011-2012.

\begin{figure}
\centering
\includegraphics[width=4.5cm,angle=90,clip=true]{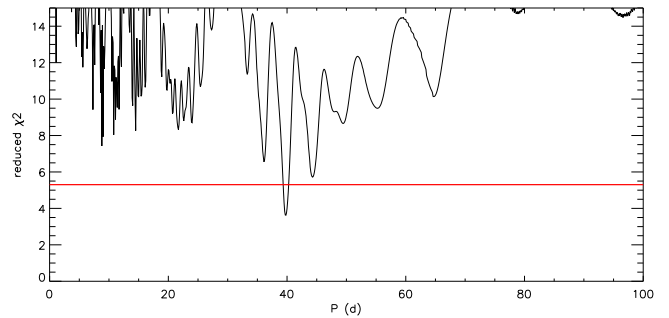}
\caption{Periodogram of the \bell\ values measured in the Stokes $V$ 2011-2012 profiles. The red horizontal line marks the 3$\sigma$ significance level above the minimum.}
\label{fig:periodo}%
\end{figure}

\begin{figure}
\centering
\includegraphics[width=4.5cm,angle=90,clip=true]{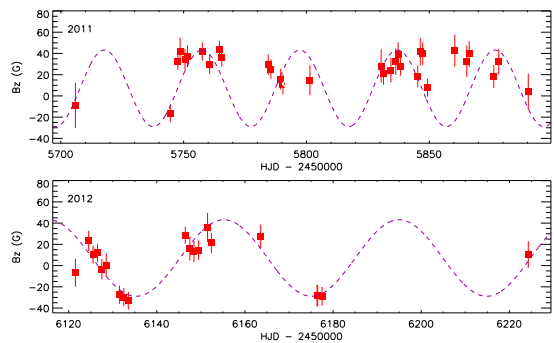}
\caption{Longitudinal magnetic field measurements (red squares) in the $V$ profiles for the 2011 (top) and 2012 (bottom) data superimposed with the best fit ($P = 39.8\pm0.5$~d, dashed lines).}
\label{fig:bell}%
\end{figure}

In order to identify a possible modulation period of the observed variations in the 2011-2012 Stokes $V$ profiles, we have computed periodograms by performing a least-squares fit of the \bell\ curve with a sinusoidal function. A sine wave is expected if the magnetic field is similar to an inclined dipole. 
%Both periodograms computed with the $N$ and $V$ 2011-2012 profiles are plotted in Fig. \ref{fig:periodo}. As no signal is expected in $N$, its periodogram helps us to determine if the minimum in the $V$ periodogram is physical or due to random noise or systematic errors. We observe that, while no well-defined minimum can be identified in the $N$ periodogram, the $V$ periodogram shows a clear minimum at $39.8\pm0.5$~d with a reduced $\chi^2$ of 3.6
The periodogram computed with the 2011-2012 \bell\ values shows a clear minimum at $39.8\pm0.5$~d with a reduced $\chi^2$ of 3.6 (Fig. \ref{fig:periodo}). The resulting best sinusoidal curve is superimposed with the measured \bell\ values in Fig. \ref{fig:bell}. The best fit reproduces well all but one of the 2012 observations, confirming that a real modulation is present in our data set. However, a number of 2011 data point are not concordant with this period. We have therefore searched for a period in each of the 2011 and 2012 data sets separately. While in the 2011 data no significant period could be identified, a period of $40\pm8$~d is clearly detected in the 2012 set, with a reduced $\chi^2$ of 1.8, consistent with the value found using the whole 2011-2012 data set. 

The relatively satisfactory fit to a sine wave in 2012, while the 2011 data fit such a wave quite poorly, together with the fact that the 2012 \bell\ data seem to vary strongly, with 8 out of 20 of the \bell\ values below 0, while only 2 in 31 of the 2011 data are below zero, suggest that the variations observed during the two epochs are quite different.

%This suggests that the 2012 data better display the periodicity than the 2011 data. However the 2011 data used in addition to the 2012 data allows us to better constrain the period. The period of $39.8\pm0.5$~d that better fits our data, if confirmed to be the rotation period of the star, would be consistent with the very low value of the projected rotation velocity ($v\sin i<8.6$~\kms) reported by \citet{catala07}.

%
%__________________________________________________________________
\section{Discussion}

From 2005 to 2009 no variation of the Zeeman signature was detected. Catala et al. (2007) proposed three hypothesis: (i) the star is seen pole-on, (ii) the rotation and magnetic axes are aligned, or (iii) the rotation period is very long (longer than the time range of our observations, i.e. several years). No fundamental changes of the photospheric line profiles (in $I$) or in the circumstellar environment have been observed from 2005 to 2012. It is therefore reasonable to assume that the rotation axis has not changed either.
If the variations of the Zeeman signature are due to rotational modulation, as suggested by the periodic variations of the 2012 data, hypothesis (i) can be ruled out. Moreover, we propose the rotation period to be about 40 days and consequently hypothesis (iii) can also be ruled out. Therefore before 2011, the magnetic axis was very likely aligned with the rotation axis (hypothesis (ii)).

It seems reasonable to assume that the modulation of the Zeeman signature observed in 2011-2012 is at least partly caused by the rotation of the star. If the magnetic configuration at the surface of the star is stable over many rotation periods, a periodicity should be observed in the Zeeman signature. In 2012 alone, all but one of the 20 \bell\ data are well reproduced with a sine wave curve, implying that a periodicity is detected. We therefore assume that the field of HD 190073 in 2012 is roughly dipolar, and inclined with respect to the rotation axis, so that the variations are due mainly to oblique rotator-like rotation. In the 2011 data, we are not able to find any periodicity using a sine wave, and the period that fits almost all the 2012 data does not fit the 2011 data at all well. The 2011 data, while not strongly variable, seem more chaotic than those from 2012. We suggest that during the 2011 observing season the field structure was changing its intrinsic geometry, and variations due to rotation were only a minor part of the observed variation. 

%However, the 2011 data help us to refine the period found in the 2012 data. This result suggests that the 2011 Zeeman signatures are also modulated with a period of $39.8 \pm 0.5$~d, but are perturbed with additional signals. We propose that, in 2011, either the magnetic configuration was not stable enough to produce a periodic signal over many rotation periods, or that the configuration of the magnetic field was more complex than a dipole. 

The spectral lines of HD 190073 are very narrow, and their broadening may be dominated by turbulence, and only slightly affected by rotation \citep{catala07}. As a consequence we may have only a limited access to the rotational Doppler resolution inside the spectral lines. This Doppler resolution is useful to study the polarisation level per rotational velocity bin, which provides strong constraints on the magnetic configuration at different longitudes and latitudes \citep[see details of Doppler imaging in e.g.][]{vogt87}. It is therefore an efficient way to detect complex fields. In the absence of rotational Doppler resolution we are unable to use this method, and both simple and complex fields can produce Zeeman signatures similar to those observed in HD 190073. 

\begin{figure}
\centering
\includegraphics[width=7cm,clip=true]{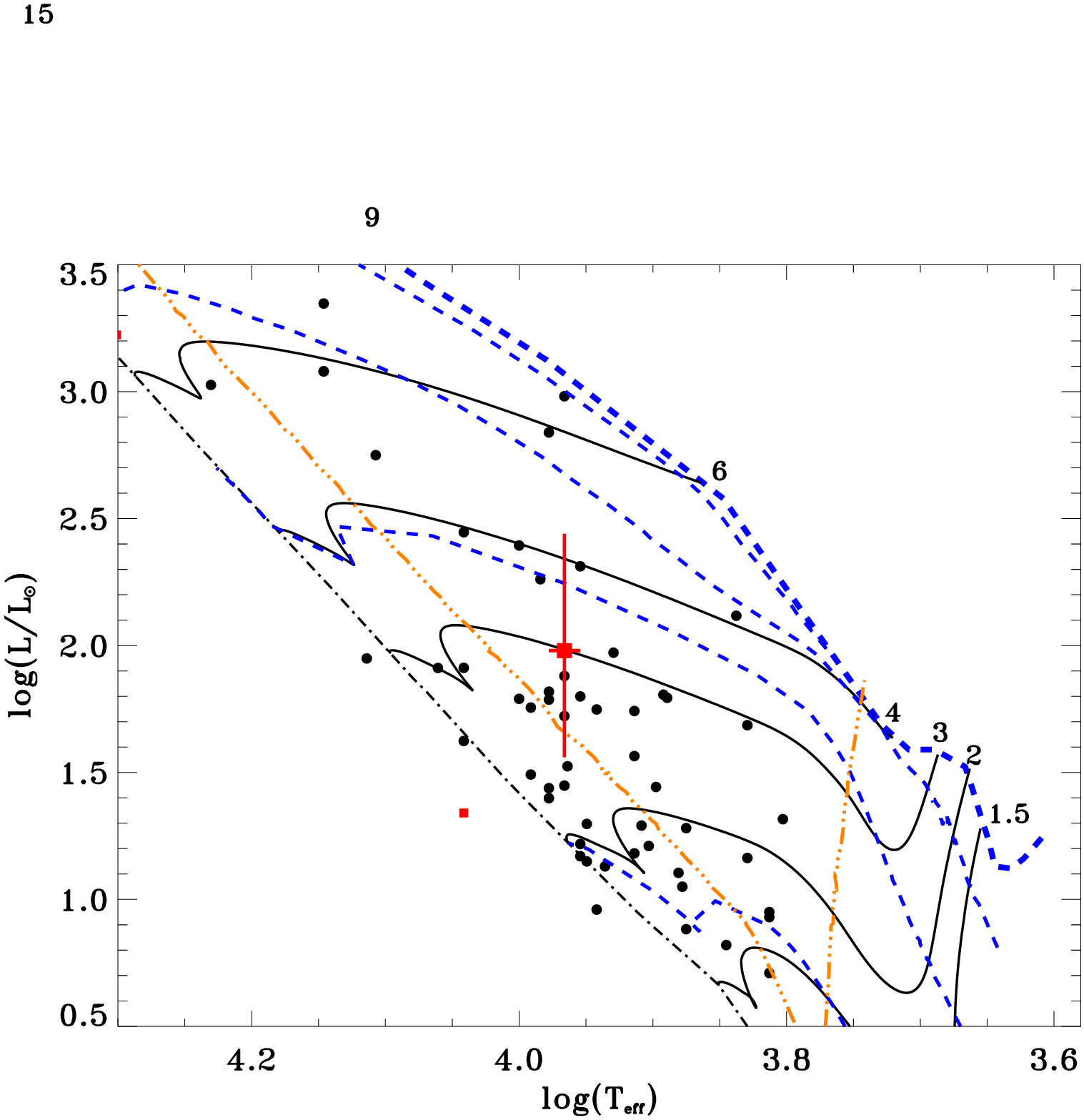}
\caption{Magnetic (red squares) and non-magnetic (black points) Herbig Ae/Be stars plotted in a Hertzsprung-Russell diagram. The red square with error bars corresponds to HD~190073. The CESAM pre-MS evolutionary tracks for 1.5, 2, 3, 4, and 6~M$_{\odot}$ (black full lines), 0.01, 0.1, 1 and 10~Myr isochrones (blue thin dashed lines), and the zero-age main-sequence (black dot-dashed line) are also plotted. The birthline taken from \citet{behrend01} is plotted with a blue thick dashed line. The limit over which the convective envelope is smaller than 1\% of the stellar mass, and the CC-birthline (see text) are overplotted with orange dot-dot-dot-dashed lines.}
\label{fig:hr}%
\end{figure}

We therefore propose the following scenario. Before 2009, the magnetic configuration of HD 190073 was stable and the magnetic axis was aligned with the rotation axis, resulting in an absence of modulation in the Zeeman signature. Between 2009 and 2011 a phenomenon occurred that perturbed the fossil field. As a consequence, the magnetic configuration lost its axisymetry, and a rotational modulation is now observed. In 2011 the magnetic field had not yet reached a stable dipolar configuration, while in 2012 the stability seems to have been achieved.

As the photosphere and circumstellar environment of HD~190073 display similar characteristics throughout this period, there is no evidence that an external event, such as interaction with a companion having an eccentric orbit, or a collision with a massive comet \citep[as observed sometimes in young Herbig stars, see e.g.][]{grinin96}, occurred during that time, and perturbed the stellar surface field. The phenomenon that perturbed the magnetic field of the star is therefore more likely of internal origin.

We have investigated the predicted internal structure of the star, by comparing its position in the HR diagram with pre-MS CESAM evolutionary tracks \citep{morel97} of solar abundance \citep{acke04b}. The distance of the system is not well known, because the Hipparcos parallax \citep[][$0.70\pm0.74$~mas]{vanleeuwen07} is very poorly constrained. Therefore we used the effective temperature of \citet[][$T_{\rm eff}=9250\pm250$~K]{catala07} and $\log g=3.75\pm0.3$ to constrain the star's position in the HR diagram. The adopted value of $\log g$ was selected because our spectra can be reproduced equally well with $\log g$ of 3.5 or 4.0 when the effective temperature is varied within the error bars. In Fig. \ref{fig:hr}, if we take into account the error bars, we observe that HD~190073 is situated just before or perhaps on the convective core birthline (CC-birthline), i.e. the condition in which the convective core has reached a mass larger than 1\% of the mass of the star.  Before reaching the CC-birthline, the star is entirely radiative and hosts a fossil field. When the star reaches the CC-birthline, and the convective core starts to build up, and convection is able to generate a dynamo magnetic field { \citep{moss89,charbonneau01,brun05}}. We expect this dynamo field to couple with the fossil field at the border with the radiative envelope. \citet{featherstone09} studied, with 3D magneto-hydrodynamical simulations, the interaction between such a convective dynamo field and a stable axisymetric twisted dipolar fossil field in a surrounding radiative envelope. They showed that the coupling induces modifications of the fossil field geometry, in particular that the field becomes oblique.

Therefore we propose that the dynamo field generated by the growing convective core has perturbed the fossil field anchored in the radiative zone of the star and is at the origin of the observed change in HD~190073. The simulations by Featherstone et al. (2009) included an already established dynamo in the core, perturbed by a fossil field in the envelope, while in HD~190073 the fossil field is present first and the convective dynamo appears progressively. Therefore it would be useful to simulate this specific case.
In addition, the observed change occurred in less than 2 years. This timescale is compatible with the proposed scenario, since stellar convective instabilities, and consequently a dynamo, grow on timescales of about 1 month \citep[e.g.][]{parker75}.

To test whether the new magnetic configuration has already stabilised and to determine the precise parameters of this configuration, additional observations will be required. HD~190073 represents a unique opportunity to witness the appearance of a convective core in a star.

\begin{acknowledgements}
S. M. thanks A.-S. Brun for fruitful discussions. We are very grateful to the TBL service observing team for its valuable help in obtaining the observations of HD 190073 at crucial times. We wish to thank the Programme National de Physique Stellaire (PNPS) for their support. This research has made use of the SIMBAD database and the VizieR catalogue access tool, operated at CDS, Strasbourg (France).
\end{acknowledgements}

\bibliographystyle{aa}
\bibliography{letter_hd190073}

\end{document}